\documentclass[aps,pre,preprint,superscriptaddress]{revtex4-2}
\usepackage{amssymb}
\usepackage{amsmath}
\usepackage{graphicx}
\usepackage{subfigure}
\usepackage{booktabs}
\usepackage{threeparttable}
\usepackage{array}
\usepackage{multirow}
\usepackage{bm}
\usepackage{CJK}

\begin{document}


\title{A multi-scale kinetic inviscid flux extracted from the gas-kinetic scheme for simulating incompressible and compressible flows}





\author{Sha Liu}
\email{shaliu@nwpu.edu.cn}
\affiliation{National Key Laboratory of Science and Technology on Aerodynamic Design and Research, Northwestern Polytechnical University, Xi'an, Shaanxi 710072, China}
\affiliation{School of Aeronautics, Northwestern Polytechnical University, Xi'an, Shaanxi 710072, China}

\author{Junzhe Cao}
\email{caojunzhe@mail.nwpu.edu.cn}
\affiliation{School of Aeronautics, Northwestern Polytechnical University, Xi'an, Shaanxi 710072, China}

\author{Chengwen Zhong}
\email{Corresponding author: zhongcw@nwpu.edu.cn}
\affiliation{National Key Laboratory of Science and Technology on Aerodynamic Design and Research, Northwestern Polytechnical University, Xi'an, Shaanxi 710072, China}
\affiliation{School of Aeronautics, Northwestern Polytechnical University, Xi'an, Shaanxi 710072, China}
\date{\today}

\begin{abstract}
A Kinetic Inviscid Flux (KIF) is proposed for simulating incompressible and compressible flows. It is constructed based on the direct modeling of multi-scale flow behaviors, which is used in the Gas-Kinetic Scheme (GKS), the Unified Gas-Kinetic Scheme (UGKS), the Discrete Unified Gas-Kinetic Scheme (DUGKS), etc.. In KIF, the discontinuities (such as the shock wave) that can not be well resolved by mesh cells are mainly solved by the Kinetic Flux Vector Splitting (KFVS) method representing the free transport mechanism (or micro-scale mechanism), while in other flow regions that are smooth, the flow behavior is solved mainly by the central-scheme-like Totally Thermalized Transport (TTT). The weights of KFVS and TTT in KIF is automatically determined by those in the theory of direct modeling. Two ways of choosing the weights in KIF are proposed, which are actually the weights adopted in the UGKS and the DUGKS, respectively. By using the test cases of Sod shock tube, rarefaction wave, the boundary layer of flat plate, the cavity flow and hypersonic flow over circular cylinder, the validity and accuracy of the present method are examined. The KIF does not suffer from the carbuncle phenomenon, and does not introduce extra numerical viscosity in smooth regions. Especially, in the case of hypersonic cylinder, it gives a quite sharp and clear density and temperature contours. The KIF can be viewed as an inviscid-viscous splitting version of the GKS. By the doing this splitting, it is easy to be used in the traditional CFD frameworks. It can also be classified as a new type in the numerical schemes based on the kinetic theory that are represented by the works in Ref.~\cite{SunA} and Ref.~\cite{ohwada2018a}, except the weights are determined by the weights of direct modeling.
\end{abstract}


\maketitle

\section{Introduction}
The numerical flux for the Euler and the Navier-Stokes equations, which are defined in the macroscopic scale to describe the nonlinear motion of fluid, is the pivot of the Computational Fluid Dynamics (CFD). In the past decades, many famous numerical fluxes are proposed, and they can be categorized into four types: the central schemes~\cite{jameson1981numerical}, the Flux Difference Splitting (FDS) type schemes~\cite{RoeApproximate}, the Flux Vector Splitting (FVS) type schemes~\cite{StegerFlux, vanleer}, the Harten-Lax-Van Leer (HLL) type schemes~\cite{goetz2017a}, and the Advection Upstream Splitting Method (AUSM) type schemes~\cite{LiouA}. The central schemes need case-dependent parameters for high speed flows in order to capture shocks~\cite{Liou2011Open}. FDS schemes often need treatments for the violation of the entropy condition and the shock anomalies for high speed flows~\cite{kim2003cures}. FVS schemes are stable for high speed flows and simple in algorithms, while they introduce extra viscoisty in the boundary layers~\cite{kitamura2010evaluation}. To combine the advantages of the FDS and the FVS schemes, HLL and AUSM type methods are proposed. Although these methods have achieved success in CFD, the numerical scheme that can predicte shock waves and boundary layer precisly, efficently, and in a robust way is still in great demand.

Besides the fluid dynamics, the schemes for macroscopic flux can also be based on the gas-kinetic theory, in which the macroscopic fluid motion and microscopic molecular motion are connected by the statistical particle distribution function. In the original works, the Equilibrium Flux Method (EFM)~\cite{PullinDirect} and Kinetic Flux Vector Splitting (KFVS) method ~\cite{Mandal1994Kinetic} were proposed independently, which are similar and can be viewed as gas-kinetic versions of FVS method. In the KFVS, the molecules that transport across the cell interface to the right are from the left cell, and vice versa. Their contribution to macroscopic flux can be obtained by summing up all the microscopic properties of molecules (In gas-kinetic theory, this calculation uses the particle distribution function). The KFVS is very robust as a shock capture scheme. While, since it has extra numerical dissipations, it is not suitable for predicting the boundary layers, so are the friction force and heat flux on solid wall. On the other hand, the Totally Thermalized Transport scheme (TTT)~\cite{TaoGas} makes the distributions at both sides of the cell interface merge into a new equilibrium (Maxwellian) distribution. It can also be viewed as a kind of central scheme. There is no extra numerical dissipation in TTT, but it is not a shock capture scheme. Since the gas-kinetic theory connects the macroscopic and microscopic flow motions, the philosophy of direct modeling was proposed formally in 2015~\cite{Xu2015Direct}, where the free-transport (corresponding to the microscopic/rarefied flow mechanism) and collision (corresponding to the macroscopic/continuum flow mechanism) of molecules are coupled together. Numerical methods based on this philosophy such as Unified Gas-Kinetic Scheme (UGKS)~\cite{xu2010unified, Li2018A, ChenA} and Discrete UGKS (DUGKS)~\cite{guo2013discrete, wang2015a, zhu2016discrete} obtain the ability of predicting flows in all flow regimes that cover both microscopic/rarefied and macroscopic/continuum flows, along with other multi-scale methods such as Gas-Kinetic Unified Algorithm (GKUA)~\cite{li2009gas, peng2016implicit} and Improved Discrete Velocity Method (IDVM)~\cite{yang2018improved, yang2019an}. Their validity and accuracy are proved by plenty of benchmark test cases, and have been extended to other physics such as plasma gas~\cite{Liu2017A}, phonon heat transfer~\cite{Guo2016Discrete} and radiative transfer~\cite{sun2017a}, etc..

Using the same coupled free-transport and collision processes of molecules and a second order C-E expansion to approximate the distribution function, the Gas-Kinetic Scheme (GKS) is constructed for simulating continuum and near-continuum flows~\cite{XuA}. The GKS method is earlier than the UGKS method and the direct modeling. Therefore, when UGKS was proposed, it was viewed as an extension of GKS to multi-scale flow predictions. From the macroscopic point of view, the inviscid flux and the viscous flux are coupled together in the GKS method. Since the GKS method is robust and accurate for flows in whole speed regime from subsonic to hypersonic~\cite{LiaoGas, LiApplication, jin2010a}, a multi-scale inviscid flux is extracted from the GKS in the paper. Therefore, the inviscid flux and viscous flux are decoupled. This treatment makes this multi-scale inviscid flux easy to be adopted in traditional CFD frameworks and makes its algorithm concise and simple. There are other methods that are based on gas-kinetic theory and combine the KFVS with other scheme for the inviscid flux calculations~\cite{SunA, Ohwada2013On}. The present method can also be classified into this type of method, except the weights are determined by the weights in the direct modeling. The remaining of this paper is organized as the following: Sec.~\ref{Sec:GKT} is a quick review of the gas-kinetic theory, Sec.~\ref{Sec:KIF} is the KIF method, the test cases are in Sec.~\ref{Sec:test}, and the conclusion is in Sec.~\ref{Sec:conclusion}.

\section{The gas-kinetic theory}\label{Sec:GKT}
In gas-kinetic theory, the physical system is described by the particle distribution function $f\left(\bm{x},\bm{\xi},\bm{\eta},t\right)$ which means the number density of particles that arrive at position $\bm{x}$ at time $t$ with velocity $\bm{\xi}$ and internal energy represented by the equivalent velocity $\bm{\eta}$. The time evolution of this distribution function is governed by the Boltzmann equation or its model equations. Besides the $\bm{x}$ and $t$ which the macroscopic primitive variables (density $\rho$, velocity $U$, $V$, $W$ and temperature $T$) depend on, $f$ also depends on extra $\bm{\xi}$ and $\bm{\eta}$.

The macroscopic conservative variables $\bm{\Psi}=\left(\rho, \rho U, \rho V, \rho W, \rho E\right)$(density, momentum density and energy density) can be directly obtained from the distribution function $f$ by the following relation:
\begin{equation}\label{eq:gkt_moment0}
\bm{{\bm{\Psi }}} = \left\langle {\bm{\varphi}f}\right\rangle,
\end{equation}
where $\bm{\varphi}=\left(m,m{\xi _1},m{\xi _2},m{\xi _3}, {{1 \over 2}m\left( {{\bm{\xi }} \cdot {\bm{\xi }} + {\bm{\eta }} \cdot {\bm{\eta }}} \right)}\right)$ is the mass, momentum and energy of a molecule. This equation (Eq.~\ref{eq:gkt_moment0}) can written explicitly as
\begin{equation}\label{eq:gkt_moment1}
\begin{aligned}
& \rho  = \left\langle {mf} \right\rangle,\\
& \rho U = \left\langle {m{\xi _1}f} \right\rangle,\\
& \rho V = \left\langle {m{\xi _2}f} \right\rangle,\\
& \rho W = \left\langle {m{\xi _3}f} \right\rangle,\\
& \rho E = {1 \over 2}\rho \left( {{U^2} + {V^2} + {W^2}} \right) + \rho I = \left\langle {{1 \over 2}m\left( {{\bm{\xi }} \cdot {\bm{\xi }} + {\bm{\eta }} \cdot {\bm{\eta }}} \right)f} \right\rangle,
\end{aligned}
\end{equation}
where
$I=\left(D+3\right)RT/2$ is the internal energy per unit mass, $R$ is specific gas constant, $D$ is the degree of freedom about the rotational and the vibrational motion of molecules; the operator $\left\langle  \cdot  \right\rangle$ is defined as the integral over the three-dimensional $\bm{\xi}$ and the $D$-dimensional $\bm{\eta}$ in the following form:
\begin{equation}\label{eq:gkt_bracket}
\left\langle  \cdot  \right\rangle {\rm{ = }}\int_{{R^3}} {d{\bm{\xi }}\int_{{R^D}} {d{\bm{\eta }}} }.
\end{equation}
The integral is called the ``moment'' in the gas-kinetic theory. The stress $\bm{S}$ and heat flux $\bm{Q}$ can also be obtained by the following moments:
\begin{equation}\label{eq:gkt_moment2}
\begin{aligned}
& {\bm{S}} = -\left\langle {m{\bm{cc}}f} \right\rangle,\\
& {\bm{Q}} = \left\langle {{\bm{c}}\left[ {{1 \over 2}m\left( {{\bm{c}} \cdot {\bm{c}} + {\bm{\eta }} \cdot {\bm{\eta }}} \right)f} \right]} \right\rangle,
\end{aligned}
\end{equation}
where $\bm{c}=\bm{\xi}-\bm{U}$ is the peculiar velocity (the macroscopic velocity $\bm{U}$ is also known as the mean velocity of all particles). For equilibrium (Maxwellian) distribution $f=g$ (Eq.~\ref{eq:maxwellian}) with the maximum entropy, both $\bm{S}$ and $\bm{Q}$ are zero. Therefore, the stress and heat flux stem from the deviation of a given $f$ from $g$.
\begin{equation}\label{eq:maxwellian}
g = {\left( {\frac{1}{{2\pi RT}}} \right)^{\frac{{D + 3}}{2}}}\exp \left( { - \frac{{{\bm{\xi }} \cdot {\bm{\xi }} + {\bm{\eta }} \cdot {\bm{\eta }}}}{{2RT}}} \right).
\end{equation}

\section{The Kinetic Inviscid Flux}\label{Sec:KIF}
The macroscopic flux $\bm{F}$ at the cell interface can be expressed by the moments of distribution as follows:
\begin{equation}\label{eq:gkt_moment_flux}
{\bm{F}} = {\left\langle {\left( {{\bm{\xi }} \cdot {\bm{n}}} \right){\bm{\varphi }}{f}} \right\rangle},
\end{equation}
where the elements of $\bm{F}$ are, in order, the fluxes of mass, momentum and energy, $\bm{n}$ is the unit normal vector of the cell interface, which points to the right side of this interface. In the remaining of Sec.~\ref{Sec:KIF}, it is assumed that $\xi_{1}$ is in the same direction of $\bm{n}$ ($\bm{\xi} \cdot \bm{n}=\xi_{1}$) without loss of generality.

\subsection{The KFVS scheme and the analysis}\label{Sec:KFVS}
In this paper, the KFVS flux (denoted by $\bm{K}$) is rewritten explicitly in the following form:
\begin{equation}\label{eq:KFVS}
\begin{aligned}
{K_{1}} =& \frac{{\rm{1}}}{{\rm{2}}}\left( {{\rho _L}{U_L} + {\rho _R}{U_R}} \right) + \frac{1}{2}\left( {{\rho _L}{U_L}{\chi _L} - {\rho _R}{U_R}{\chi _R}} \right) + \frac{1}{2}\left( {{\rho _L}{\theta _L} - {\rho _R}{\theta _R}} \right),\\
{K_{2}} =& \frac{{\rm{1}}}{{\rm{2}}}\left[ {\left( {{\rho _L}U_L^2 + {P_L}} \right) + \left( {{\rho _R}U_R^2 + {P_R}} \right)} \right] \\
&+ \frac{{\rm{1}}}{{\rm{2}}}\left[ {\left( {{\rho _L}U_L^2 + {P_L}} \right){\chi _L} - \left( {{\rho _R}U_R^2 + {P_R}} \right){\chi _R}} \right] + \frac{{\rm{1}}}{{\rm{2}}}\left( {{\rho _L}{U_L}{\theta _L} - {\rho _R}{U_R}{\theta _R}} \right),\\
{K_{3}} =& \frac{{\rm{1}}}{{\rm{2}}}\left( {{\rho _L}{U_L}{V_L} + {\rho _R}{U_R}{V_R}} \right) + \frac{1}{2}\left( {{\rho _L}{U_L}{V_L}{\chi _L} - {\rho _R}{U_R}{V_R}{\chi _R}} \right) + \frac{1}{2}\left( {{\rho _L}{V_L}{\theta _L} - {\rho _R}{V_R}{\theta _R}} \right),\\
{K_{4}} =& \frac{{\rm{1}}}{{\rm{2}}}\left( {{\rho _L}{U_L}{W_L} + {\rho _R}{U_R}{W_R}} \right) + \frac{1}{2}\left( {{\rho _L}{U_L}{W_L}{\chi _L} - {\rho _R}{U_R}{W_R}{\chi _R}} \right) + \frac{1}{2}\left( {{\rho _L}{W_L}{\theta _L} - {\rho _R}{W_R}{\theta _R}} \right),\\
{K_{5}} =& \frac{{\rm{1}}}{{\rm{2}}}\left( {{\rho _L}{U_L}{H_L} + {\rho _R}{U_R}{H_R}} \right)\\
&+ \frac{1}{2}\left( {{\rho _L}{U_L}{H_L}{\chi _L} - {\rho _R}{U_R}{H_R}{\chi _R}} \right) + \frac{1}{2}\left[ {\left( {{\rho _L}{H_L} - \frac{{{P_L}}}{2}} \right){\theta _L} - \left( {{\rho _R}{H_R} - \frac{{{P_R}}}{2}} \right){\theta _R}} \right],
\end{aligned}
\end{equation}
where $H=E+P/\rho$ ($P$ is the hydrodynamic pressure) is the enthalpy per unit mass, the subscript ``L" and ``R" represent the left and right hand sides of the cell interface. $\chi$ and $\theta$ are defined as:
\begin{equation}\label{eq:coeff_KFVS}
\begin{aligned}
&{\chi _\alpha } = {\rm{erf}}\left( {\frac{{{U_\alpha }}}{{\sqrt {2R{T_\alpha }} }}} \right),\\
&{\theta _\alpha } = \sqrt {\frac{{2R{T_\alpha }}}{\pi }} \exp \left( { - \frac{{U_\alpha ^2}}{{2R{T_\alpha }}}} \right),
\end{aligned}
\end{equation}
where the subscript ``$\alpha$" can be ``L" or ``R". Also, $\chi$ and $\theta$ can be expressed in words of the Mach number (Ma) and sonic speed ($a$) as:
\begin{equation}\label{eq:coeff_KFVS2}
\begin{aligned}
& {\chi _\alpha } = {\rm{erf}}\left( {\sqrt {{\gamma  \over 2}} {\rm{M}}{{\rm{a}}_\alpha }} \right),\\
& {\theta _\alpha } = {a_\alpha }\sqrt {{2 \over {\gamma \pi }}} \exp \left( { - {\gamma  \over 2}{\rm{Ma}}_\alpha ^2} \right),
\end{aligned}
\end{equation}
where $\gamma=\left(D+5\right)/\left(D+3\right)$ is the specific heat ratio.

The above is the scheme of KFVS. It is derived from the following moment equation for flux,
\begin{equation}\label{eq:KFVS_moment}
{\bm{F}} = {\left\langle {\left( {{\bm{\xi }} \cdot {\bm{n}}} \right){\bm{\varphi }}{g_L}} \right\rangle _R} + {\left\langle {\left( {{\bm{\xi }} \cdot {\bm{n}}} \right){\bm{\varphi }}{g_R}} \right\rangle _L},
\end{equation}
where the half moments ${\left\langle  \cdot  \right\rangle _L}$ and ${\left\langle  \cdot  \right\rangle _R}$ are defined as
\begin{equation}
\begin{aligned}
{\left\langle  \cdot  \right\rangle _L}=\int_{ - \infty }^0 {d{\xi _1}\int_{ - \infty }^{ + \infty } {d{\xi _2}\int_{ - \infty }^{ + \infty } {d{\xi _3}\int_{{R^D}} {\left(  \cdot  \right)d{\bm{\eta }}} } } },\\
{\left\langle  \cdot  \right\rangle _R}=\int_0^{ + \infty } {d{\xi _1}\int_{ - \infty }^{ + \infty } {d{\xi _2}\int_{ - \infty }^{ + \infty } {d{\xi _3}\int_{{R^D}} {\left(  \cdot  \right)d{\bm{\eta }}} } } }.
\end{aligned}
\end{equation}
The calculation of these half moments is not in the scope of this paper, and interested readers can refer to Ref.~\cite{XuA}.

This moment equation for flux is actually Eq.~\ref{eq:gkt_moment_flux} with the general distribution $f$ being replaced by two half Maxwellian distributions ($g_L$ and $g_R$),
\begin{equation}\label{Eq:KFVS_distri}
f = \left\{ {\begin{array}{*{20}{c}}
{{g_L},{\bm{\xi }} \cdot {\bm{n}} \ge 0}\\
{{g_R},{\bm{\xi }} \cdot {\bm{n}} < 0}
\end{array}} \right..
\end{equation}
The physical meaning of Eq.~\ref{eq:KFVS_moment} is that, at the cell interface, $f$ is approximated by two half Maxwellian distributions ($g_{L}$ and $g_{R}$) which are determined by the macroscopic variables on the corresponding sides ($\bm{\Psi}_{L}$ and $\bm{\Psi}_{R}$) through Eq.~\ref{eq:maxwellian}; the particles toward left (corresponding to the operator ${\left\langle  \cdot  \right\rangle _L}$) are from the right hand side (corresponding to $g_{R}$), and vice versa.

Like the FVS, the KFVS has an extra viscosity in the numerical scheme. From the perspective of the gas-kinetic theory, when the two parts of the distributions (half of $g_L$ and half of $g_R$) are combined together, the resultant distribution function (Eq.~\ref{Eq:KFVS_distri}) is not a Maxwellian one, unless the values of $\bm{\Psi}_{L}$ and $\bm{\Psi}_{R}$ are the same. But, the reconstruction process of the macroscopic conservative variables often leads to the different $\bm{\Psi}_{L}$ and $\bm{\Psi}_{R}$. Therefore, the deviation of the distribution function from the Maxwellian one leads to the extra stress and heat flux (see Sec.~\ref{Sec:GKT}). These extra stress and heat flux make the KFVS be diagnosed with an extra and unphysical viscosity. Here, it is clear that implicit reason of the extra viscosity is the deviation of the distribution function from the Maxwellian one.

\subsection{The TTT scheme and the analysis}\label{Sec:TTT}
Given that the flux should be calculated from a Maxwellian distribution without any deviation from it, if one wants to construct a scheme without any extra viscosity in smooth regions of the flow field. The two half distributions ($g_{L}$ and $g_{R}$ in Eq.~\ref{Eq:KFVS_distri}) are merged into a new single Maxwellian distribution through the following process in TTT:
\begin{equation}\label{eq:TTT_basic}
{\bm{\bar \Psi}} = {\left\langle {{\bm{\varphi }}{g_L}} \right\rangle _L} + {\left\langle {{\bm{\varphi }}{g_R}} \right\rangle _R} = {\left\langle {{\bm{\varphi }}{\bar g}} \right\rangle}.
\end{equation}
The horizontal bar is used to represent the merged macroscopic variables and distribution function. The merged conservative variables can be explicitly obtained (from Eq.~\ref{eq:TTT_basic}) as follows:
\begin{equation}\label{eq:TTT}
\begin{aligned}
&\overline \rho   = \frac{{\rm{1}}}{{\rm{2}}}\left( {{\rho _L}{\rm{ + }}{\rho _R}} \right) + \frac{{\rm{1}}}{{\rm{2}}}\left( {{\rho _L}{\chi _L} - {\rho _R}{\chi _R}} \right),\\
&\overline {\rho u}  = \frac{{\rm{1}}}{{\rm{2}}}\left( {{\rho _L}{U_L}{\rm{ + }}{\rho _R}{U_R}} \right) + \frac{1}{2}\left( {{\rho _L}{\theta _L} - {\rho _{\rm{R}}}{\theta _R}} \right) + \frac{1}{2}\left( {{\rho _L}{U_L}{\chi _L} - {\rho _R}{U_R}{\chi _R}} \right),\\
&\overline {\rho v}  = \frac{{\rm{1}}}{{\rm{2}}}\left( {{\rho _L}{V_L}{\rm{ + }}{\rho _R}{V_R}} \right) + \frac{{\rm{1}}}{{\rm{2}}}\left( {{\rho _L}{V_L}{\chi _L} - {\rho _R}{V_R}{\chi _R}} \right),\\
&\overline {\rho w}  = \frac{{\rm{1}}}{{\rm{2}}}\left( {{\rho _L}{W_L}{\rm{ + }}{\rho _R}{W_R}} \right) + \frac{{\rm{1}}}{{\rm{2}}}\left( {{\rho _L}{W_L}{\chi _L} - {\rho _R}{W_R}{\chi _R}} \right),\\
&\overline {\rho E}  = \frac{{\rm{1}}}{{\rm{2}}}\left( {{\rho _L}{E_L}{\rm{ + }}{\rho _R}{E_R}} \right) + \frac{1}{4}\left( {{\rho _L}{U_L}{\theta _L} - {\rho _{\rm{R}}}{U_R}{\theta _R}} \right) + \frac{1}{2}\left( {{\rho _L}{E_L}{\chi _L} - {\rho _R}{E_R}{\chi _R}} \right),
\end{aligned}
\end{equation}
where $\chi$ and $\theta$ are defined in Eq.~\ref{eq:coeff_KFVS2}. Then the TTT flux (denoted by $\mathbf{G}$) is the simple Euler flux using the merged macroscopic variables as follows:
\begin{equation}\label{eq:Euler}
\begin{aligned}
&{G_{1}} = \bar \rho \bar U,\\
&{G_{2}} = \bar \rho {{\bar U}^2} + \bar P,\\
&{G_{3}} = \bar \rho \bar U\bar V,\\
&{G_{4}} = \bar \rho \bar U\bar W,\\
&{G_{5}} = \bar \rho \bar U\bar H,
\end{aligned}
\end{equation}
where $\bar U=\overline {\rho U}/\bar \rho$, $\bar V=\overline {\rho V}/\bar \rho$, $\bar W=\overline {\rho W}/\bar \rho$, $\bar H=\gamma \overline{\rho E}/\bar \rho$ and $\bar P = \left(\gamma-1\right)\overline{\rho E}$.

Since the TTT flux is totally from the exact Maxwellian distribution, there is no extra viscosity in TTT at all. As a result, TTT is not a shock capture scheme and is not stable at the discontinuity.

\subsection{The KIF scheme and the analysis}
\subsubsection{The KIF scheme}
In the section, it is organized that: first, the KIF scheme is proposed explicitly, along with its numerical procedure; then, its physical background (from gas-kinetic theory and GKS) is analyzed.

The KIF flux is
\begin{equation}\label{Eq:KIF}
\bm{F}_{KIF}=\beta \bm{K}+\left(1-\beta\right)\bm{G},
\end{equation}
where $\bm{K}$ and $\bm{G}$ are the KFVS flux (Eq.~\ref{eq:KFVS}) and TTT flux (Eq.~\ref{eq:Euler}), respectively; $\beta$ is a weight factor depended on the multi-scale mechanism of the flow around discontinuities. Therefore, Eq.~\ref{Eq:KIF} can be view as a weighted average of of KFVS and TTT fluxes.

Two types of $\beta$ are proposed in this paper. The corresponding flux are denoted by KIF1 and KIF2, respectively.

For KIF1, $\beta$ is defined as
\begin{equation}\label{Eq:beta1}
\beta =\frac{{1 - {e^{ - r}}}}{r}.
\end{equation}
For KIF2, it is
\begin{equation}\label{Eq:beta2}
\beta  = \frac{1}{{1 + \dfrac{r}{2}}}.
\end{equation}
In both Eq.~\ref{Eq:beta1} and Eq.~\ref{Eq:beta2}, the variable $r$ is defined as
\begin{equation}\label{Eq:ratio}
r = \frac{1}{\mathop {\max }\limits_{\omega  \in \Omega } {\left( {\frac{{\left| {{P_L} - {P_R}} \right|}}{{\left| {{P_L} + {P_R}} \right|}}\max \left( {{\rm{M}}{{\rm{a}}_L},{\rm{M}}{{\rm{a}}_R}} \right)} \right)_\omega}},
\end{equation}
where $\Omega$ is a set of cell interfaces that includes the target interface and the other interfaces of its adjacent cells (left and right neighbor cells).

\subsubsection{The multi-scale physical basis of KIF}
It is widely accepted that a well-designed central scheme can handle the smooth flow without extra numerical viscosity, while it is unstable when discontinuity occurs. Therefore, numerical dissipation should be directly added into it, or alternative methods should be used.

Since the scale of discontinuity is measured in the mean free path of molecules, it can not be well resolved by the resolutions of (macroscopic) mesh cells. Therefore, it is beneficial to take the microscopic flow mechanism of the discontinuity into consideration when constructing the flux scheme. Taking the shock wave discontinuity for example, its profile is actually smooth in micro-scale (zoom into the inner of the shock wave). Since the scale of shock wave is that of the mean free path (the thickness of a normal shock wave is about twenty times of mean free path) , it is reasonable to investigate it in a time scale of molecular mean collision time. Within the mean collision time, a portion of molecules are free transport ones without collision, the other molecules experience one or more collisions.

Recall that the mechanism in KFVS (Sec.~\ref{Sec:KFVS}) is that: first, the distribution functions on both sides of the cell interface are approximated by the Maxwellian distributions; then, the molecules that are on the left side and move toward right (corresponding to $g_L$ and $\left\langle  \cdot  \right\rangle _{R}$) will transport across this interface and contribute to the flux, and vice versa. This mechanism of KFVS implies that the molecules do not experience any collision when they transport across the interface, since the rightward transporting molecules strictly follow the distribution function $g_{L}$ on the left side (any collision with the $g_R$ molecules will change their distribution), and vise versa. Therefore, the KFVS actually describes the behaviors of free transport molecules.

Also recall the mechanism in TTT (Sec.~\ref{Sec:TTT}), the right half of $g_{L}$ and the left half of $g_R$ are merged into a single Maxwellian distribution $\bar g$. The merging process is actually the collision process of these molecules. The implied mechanism is that: initially, the two parts of distribution when combining together, deviate from the equilibrium Maxwellian distribution function, and stress and heat flux are generated; after sufficient collision, both stress and heat flux are decayed into zero, and $\bar g$ is achieved; since the Maxwellian distribution has the locally maximum entropy, $\bar g$ can be viewed as being totally thermalized. This merging process can finish during about 10 molecular mean collision time~\cite{Liu2019Conservative, liu2014investigation}. Therefore, the TTT actually describes the behaviors of molecules that experience sufficient collisions.

Now, it is clear that if the flux scheme wants to describe the mechanism in discontinuity (such as the shock wave), both the free transport molecules and the colliding molecules should be considered. Therefore, the KIF combines the KFVS and TTT together. Their weights are determined by the direct modeling, specifically speaking: using the particular weights of the UGKS(or GKS) and DUGKS methods.

To derive the weights used in KIF, the following BGK model equation (Eq.~\ref{Eq:BGK}) should be used,
\begin{equation}\label{Eq:BGK}
\frac{{\partial f}}{{\partial t}} + {\bm{\xi }} \cdot \frac{{\partial f}}{{\partial {\bm{x}}}} = \frac{{g - f}}{\tau },
\end{equation}
where $\tau={\mu}/{P}$ is the relaxation time whose magnitude is the same order of the mean collision time, and $\mu$ is the dynamic viscosity. Since it is a quasi-linear equation (it is linear when supposing that $g$ is a constant), an analytical solution can be obtained. For clarity and simplicity, the equilibrium distribution $g$ at the cell interface is supposed to be a constant in space and in a time interval (from the perspective of numerical methods, both space and time resolutions are the first order), and the initial distributions denoted by $f_{L}\left(0\right)$ and $f_{R}\left(0\right)$ are also piece-wise constants. Focusing on the cell interface, the analytical (time integral) solution can be explicitly written as
\begin{equation}\label{Eq:analytical}
f\left( t \right) = \left\{ {\begin{array}{*{20}{c}}
{{e^{ - \frac{t}{\tau }}}{f_L}\left( 0 \right) + \left( {1 - {e^{ - \frac{t}{\tau }}}} \right)g,{\xi _1} \ge 0}\\
{{e^{ - \frac{t}{\tau }}}{f_R}\left( 0 \right) + \left( {1 - {e^{ - \frac{t}{\tau }}}} \right)g,{\xi _1} < 0}
\end{array}} \right.,
\end{equation}
where $f\left( t \right)$ is the distribution function at time $t$ at the cell interface, and it is a weighted average of $f\left( 0 \right)$ and $g$. For both directions of molecular velocity $\xi_{1}$, the first term is the free transport term and the second term is the equilibrium term after sufficient molecular collisions. To calculate the flux, $f\left( t \right)$ is further multiplied by $\xi_{1}\bm{\varphi}$ and integrated in time interval $\left(0,\delta t\right)$, and the result is
\begin{equation}\label{eq:KIF_flux_basic}
\begin{aligned}
{\bm{F}} &= \frac{1}{{\delta t}}\int_0^{\delta t} {\left\langle {{\xi _1}{\bm{\varphi }}f\left( t \right)} \right\rangle } dt\\
&= \left\{ {\frac{\tau }{{\delta t}}\left( {1 - {e^{ - {{\delta t} \mathord{\left/
 {\vphantom {{\delta t} \tau }} \right.
 \kern-\nulldelimiterspace} \tau }}}} \right)} \right\}\left[ {{{\left\langle {{\xi _1}{\bm{\varphi }}{f_L}\left( 0 \right)} \right\rangle }_R}{\rm{ + }}{{\left\langle {{\xi _1}{\bm{\varphi }}{f_R}\left( 0 \right)} \right\rangle }_L}} \right]{\rm{ + }}\left\{ {1 - \frac{\tau }{{\delta t}}\left( {1 - {e^{ - {{\delta t} \mathord{\left/
 {\vphantom {{\delta t} \tau }} \right.
 \kern-\nulldelimiterspace} \tau }}}} \right)} \right\}\left\langle {{\xi _1}{\bm{\varphi }}g} \right\rangle.
\end{aligned}
\end{equation}

Since $g$ is the equilibrium distribution after sufficient collisions, it is reasonable to be replaced by $\bar g$ of TTT. If $f_{L}\left(0\right)$ and $f_{R}\left(0\right)$ are further replaced by $g_{L}\left(0\right)$ and $g_{R}\left(0\right)$, by given Eq.~\ref{eq:KFVS} and Eq.~\ref{eq:Euler}, this flux becomes
\begin{equation}\label{eq:KIF_flux_basic1}
{\bm{F}} = \left\{ {\frac{\tau }{{\delta t}}\left( {1 - {e^{ - {{\delta t} \mathord{\left/
{\vphantom {{\delta t} \tau }} \right.
\kern-\nulldelimiterspace} \tau }}}} \right)} \right\}{\bm{K}}{\rm{ + }}\left\{ {1 - \frac{\tau }{{\delta t}}\left( {1 - {e^{ - {{\delta t} \mathord{\left/
{\vphantom {{\delta t} \tau }} \right.
\kern-\nulldelimiterspace} \tau }}}} \right)} \right\}{\bm{G}}.
\end{equation}
$\delta t$ is the observation time scale and is measured in mean collision time (or the relaxation time $\tau$) in discontinuities, it can be set as
\begin{equation}\label{eq:deltat}
\delta t={r}{\tau}=\frac{\tau }{{\mathop {\max }\limits_{\omega  \in \Omega } {{\left[ {\frac{{\left| {{P_L} - {P_R}} \right|}}{{\left| {{P_L} + {P_R}} \right|}}\max \left( {{\rm{M}}{{\rm{a}}_L},{\rm{M}}{{\rm{a}}_R}} \right)} \right]}_\omega }}},
\end{equation}
where the $\delta t$ is relatively small when the discontinuity is severe. Submitting Eq.~\ref{eq:deltat} into Eq.~\ref{eq:KIF_flux_basic1}, the flux of KIF1 (Eq.~\ref{Eq:KIF}) is directly obtained.

The construction of KIF2 is similar, except the following characteristic-line solution~\cite{guo2013discrete} is used instead of the analytical solution in Eq.~\ref{Eq:analytical},
\begin{equation}\label{Eq:characterline}
f\left( t \right) = \left\{ {\begin{array}{*{20}{c}}
{\left( {\frac{\tau }{{\tau  + h}}} \right){f_L}\left( 0 \right) + \left( {\frac{h}{{\tau  + h}}} \right)g,{\xi _1} \ge 0}\\
{\left( {\frac{\tau }{{\tau  + h}}} \right){f_R}\left( 0 \right) + \left( {\frac{h}{{\tau  + h}}} \right)g,{\xi _1} < 0}
\end{array}} \right.,
\end{equation}
where $h=\delta t/2$. $f\left( h \right)$ is the distribution function at the cell interface at half observation time. Submit Eq.~\ref{Eq:characterline} into the definition of flux (Eq.~\ref{eq:gkt_moment_flux}), replace $f_{L}\left(0\right)$, $f_{R}\left(0\right)$ and $g$ by $g_{L}\left(0\right)$, $g_{R}\left(0\right)$ and $\bar g$, respectively, and use the $\delta t$ in Eq.~\ref{eq:deltat}, the flux of KIF2 (Eq.~\ref{Eq:KIF}) is directly obtained.

The above is the construction process and physical basis of the KIF scheme. It should be noticed that the observation time $\delta t$ in Eq.~\ref{eq:deltat} is different from the iteration time step $\Delta t$ of numerical methods. $\delta t$ can be even smaller than mean collision time in discontinuities, and makes KFVS dominate the scheme; also, it can be in the macroscopic time scale when the flow properties change smoothly across cell interfaces, and makes TTT dominate the scheme. It should also be noticed that the present KIF method can be viewed as an inviscid version of the GKS method in Ref.~\cite{XuA} with a first order model accuracy.

\section{The test cases}\label{Sec:test}
\subsection{The sod case}
The modified version of Sod case in Ref.~\cite{Toro2009Riemann} is used in this section. The initial condition of the sod case is shown below, where the flow field is divided into two parts
\begin{equation}
\begin{aligned}
&{\left( {{\rho _L},{U_L},{P_L}} \right) = \left( {1.0,0.75,1} \right),0 < x < 0.3}\\
&{\left( {{\rho _R},{U_R},{P_R}} \right) = \left( {0.125,0,0.1} \right),0.3 < x < 1}.
\end{aligned}
\end{equation}
The evolution of this flow field includes a right-traveling contact wave, a right-traveling shock wave and a rarefaction wave on the left of flow field. The case is used to test whether a scheme fulfills the entropy condition and has the capability of capturing shock wave and contact-discontinuity.

The whole flow field is divided into 100 cells. The CFL number is set 0.67, and the time step is 0.0025. The second order MUSCL is used for spatial reconstruction with Venkatakrishnan limiter~\cite{venkatakrishnan1995convergence}, and third order Runge-Kutta method is used for time advancing. The Dirichlet boundary condition is used for two boundaries whose values are set their initial values.

\begin{figure}
\centering
\subfigure[\label{Fig:sod_density} Density]{
\includegraphics[width=0.45\textwidth]{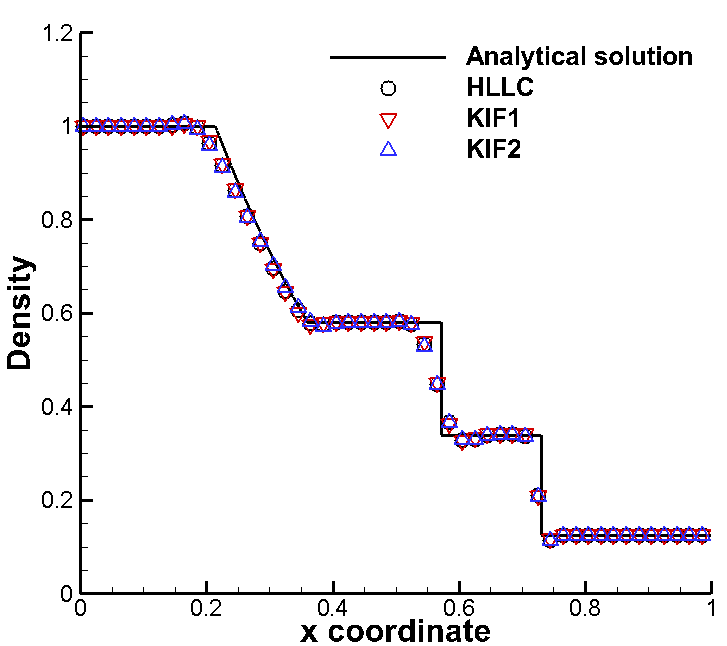}
}\hspace{0.05\textwidth}%
\subfigure[\label{Fig:sod_velocity} Velocity]{
\includegraphics[width=0.45\textwidth]{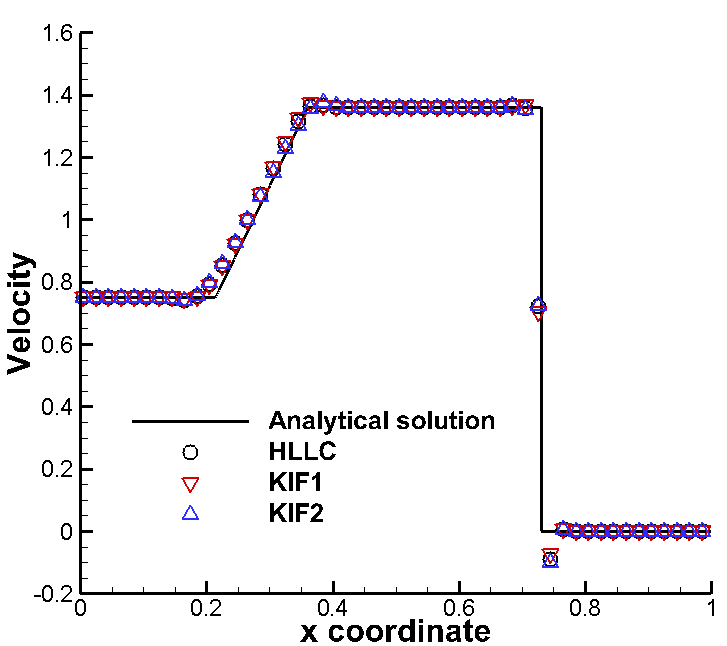}
}\\
\subfigure[\label{Fig:sod_pressure} Pressure]{
\includegraphics[width=0.45\textwidth]{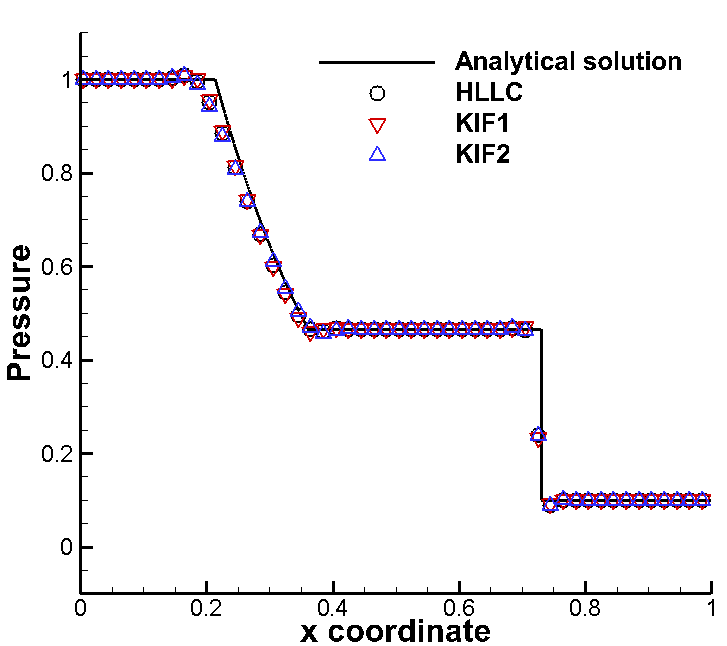}
}\hspace{0.05\textwidth}%
\subfigure[\label{Fig:sod_energy} Internal energy]{
\includegraphics[width=0.45\textwidth]{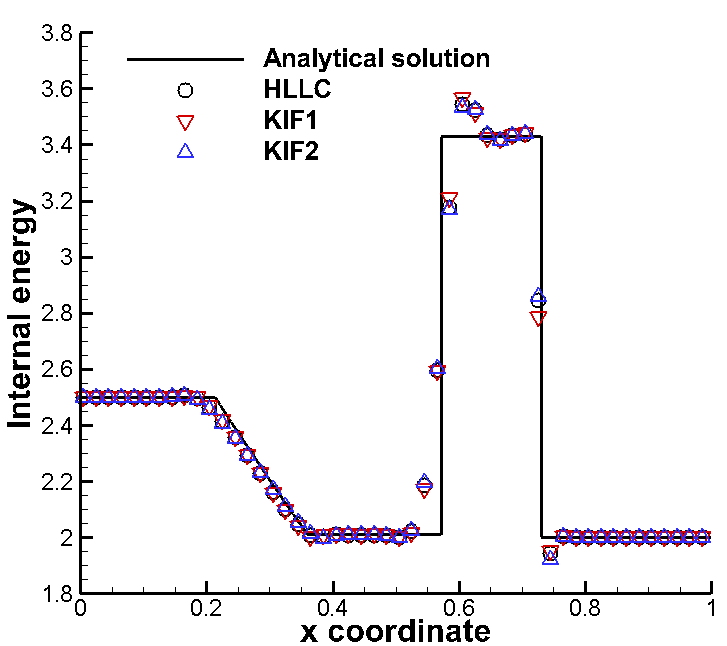}
}
\caption{\label{Fig:sod} The profiles of flow variables of Sod shock-tube at $t=0.2$}
\end{figure}

The flow field at $t=0.2$ is illustrated in Fig.~\ref{Fig:sod} and compared with the analytical solution of Euler equations and the HLLC numerical solution~\cite{ToroRestoration} (since the reconstruction affects the numerical results. The code of HLLC is used in the present framework in order to get close investigation). Both KIF1 and KIF2 match well with the analytical solution and the HLLC, and their results almost coincide with each other.

\subsection{The rarefaction wave}
The initial condition of the rarefaction wave is as follows
\begin{equation}
\begin{aligned}
{\left( {{\rho _L},{U_L},{P_L}} \right) = \left( {1, - 2,0.4} \right),0 < x < 0.5}\\
{\left( {{\rho _R},{U_R},{P_R}} \right) = \left( {1,2,0.4} \right),0.5 < x < 1},
\end{aligned}
\end{equation}
where the two parts of the flow are traveling in the opposite directions and their interface is a contact discontinuity. The two rarefaction waves extract the central regime, making the central point a near vacuum state. This case is used to test the positivity of numerical schemes and the ability of capturing the discontinuity.

The whole flow field is divided into 100 cells. The CFL number is set 0.67, and the time step is 0.0025. The second order MUSCL is used with Venkatakrishnan limiter, the third order Runge-Kutta method and Dirichlet boundary condition are also used in this case.

\begin{figure}
\centering
\subfigure[\label{Fig:rare_density} Density]{
\includegraphics[width=0.45\textwidth]{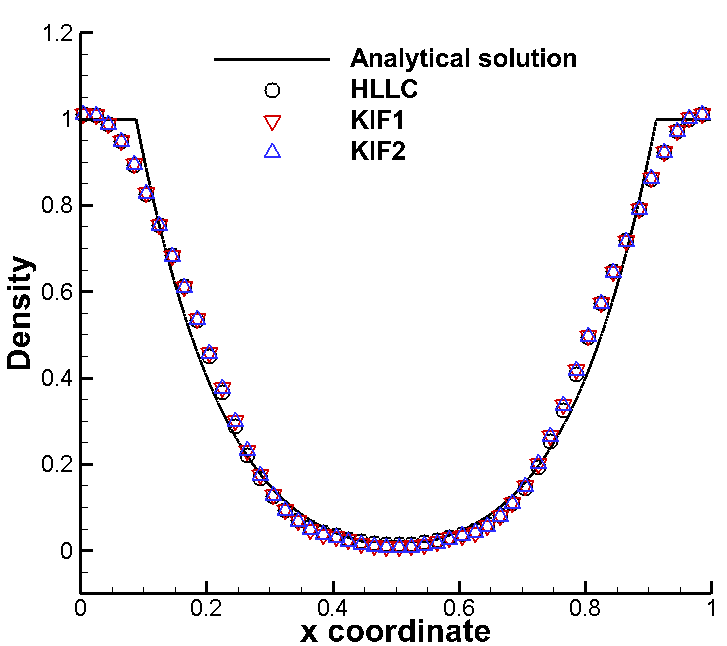}
}\hspace{0.05\textwidth}%
\subfigure[\label{Fig:rare_velocity} Velocity]{
\includegraphics[width=0.45\textwidth]{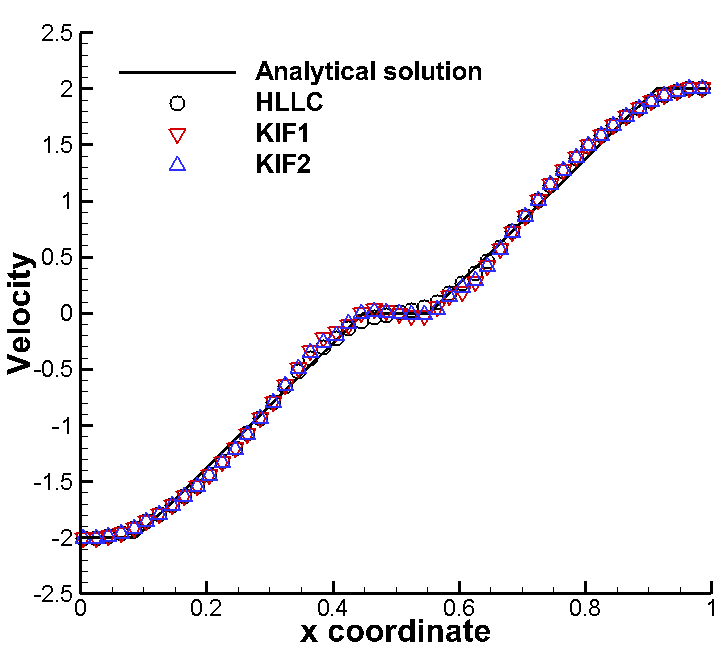}
}\\
\subfigure[\label{Fig:rare_pressure} Pressure]{
\includegraphics[width=0.45\textwidth]{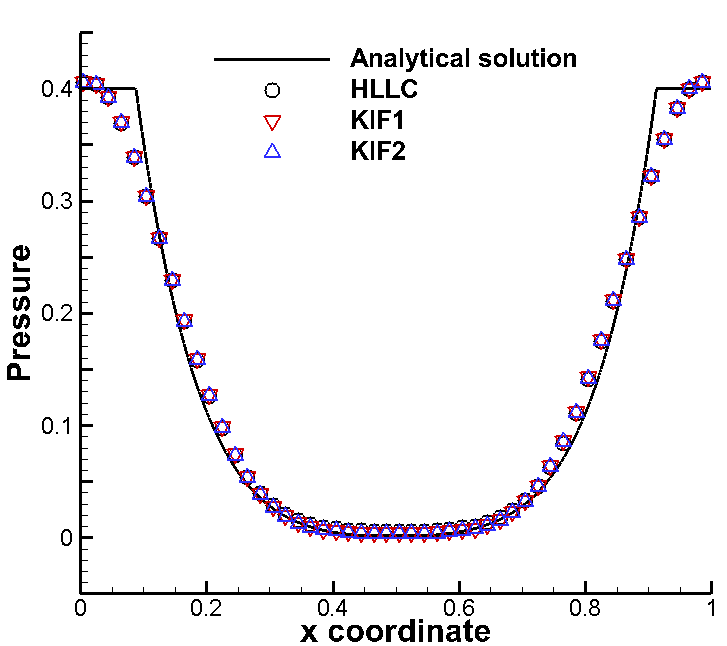}
}\hspace{0.05\textwidth}%
\subfigure[\label{Fig:rare_energy} Internal energy]{
\includegraphics[width=0.45\textwidth]{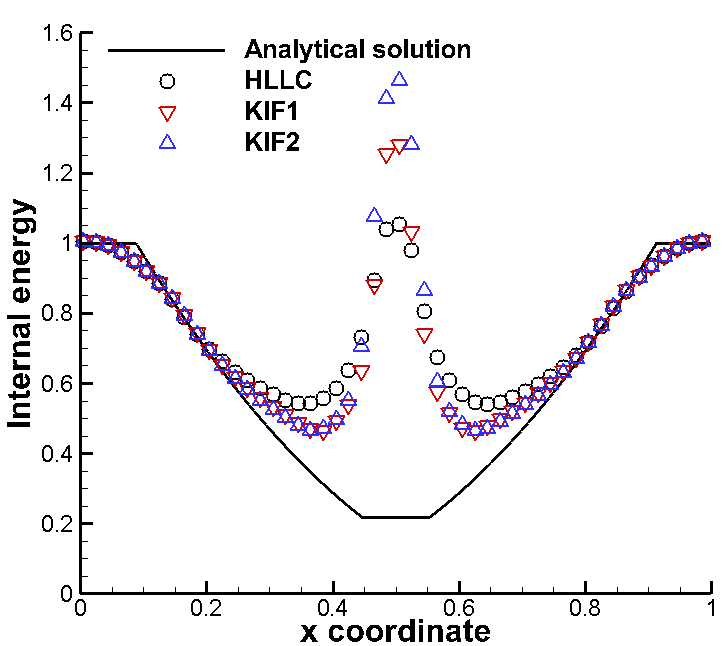}
}
\caption{\label{Fig:rare} The profiles of flow variables of the rarefaction waves at $t=0.15$}
\end{figure}

The flow field at $t=0.15$ is illustrated in Fig.~\ref{Fig:rare} and compared with the analytical solution of Euler equations and the HLLC numerical solution~\cite{ToroRestoration}.

In the density, velocity and pressure profiles, both KIF1 and KIF2 match well with HLLC results. Since the density is very low around the central regime, it makes the internal energy very sensitive. The KIF1 and KIF2 are better than HLLC in the rarefaction wave, while HLLC is better in the vacuum regime.

\subsection{The boundary layer}
In this case, the flow passes over a flat plate, and boundary layer is formed because of the viscous effect. The Blasius solution is used as the benchmark solution. If a scheme introduces extra numerical viscosity, it will deviates from the Blasius solution.

\begin{figure}
\centering
\includegraphics[width=0.45\textwidth]{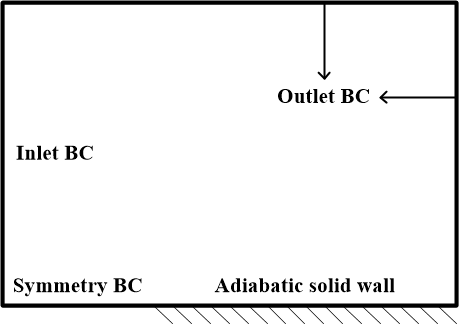}
\caption{\label{Fig:plate} The geometry of the flow over flat plate}
\end{figure}

The flow field is a rectangular area (the geometry with boundary condition is shown in Fig.~\ref{Fig:plate}), where the left surface is set as the subsonic inlet boundary(total temperature and total pressure are fixed), the top surface and right surface are set as subsonic outlet boundaries(static pressure is fixed), the plate is set as adiabatic boundary, the part of bottom surface before the plate is set as the symmetric boundary.

The inflow Ma number is $0.1$, and the inflow Re number is $1\times10^5$ (the reference length is the plate length 100). The smallest mesh is around the leading edge of plate, where $\Delta x=0.1$ and $\Delta y=0.02$ are chosen. The CFL number is set 0.8.

\begin{figure}
\centering
\subfigure[\label{Fig:plate_contour_rho} Density]{
\includegraphics[width=0.45\textwidth]{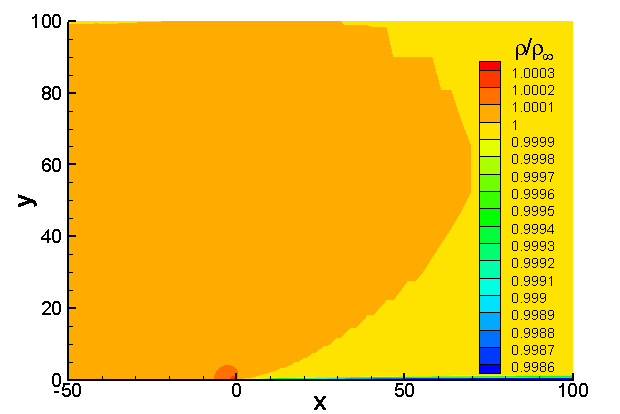}
}\hspace{0.05\textwidth}%
\subfigure[\label{Fig:plate_contour_u} U-velocity]{
\includegraphics[width=0.45\textwidth]{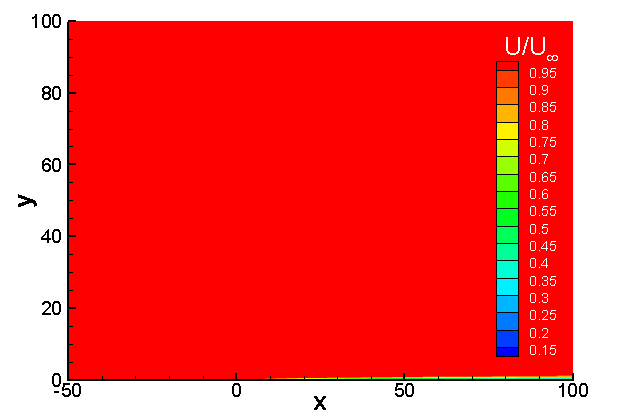}
}\\
\subfigure[\label{Fig:plate_contour_v} V-velocity]{
\includegraphics[width=0.45\textwidth]{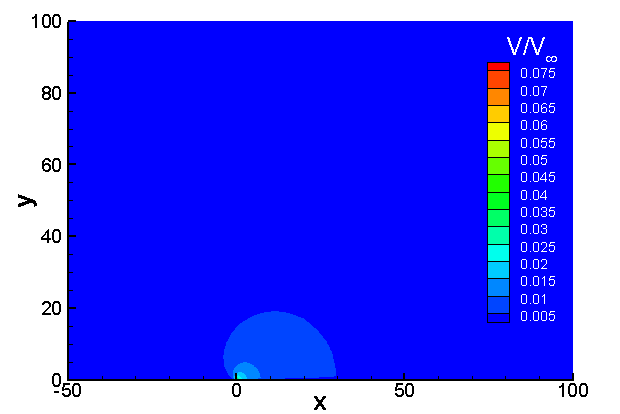}
}\hspace{0.05\textwidth}%
\subfigure[\label{Fig:plate_contour_t} Temperature]{
\includegraphics[width=0.45\textwidth]{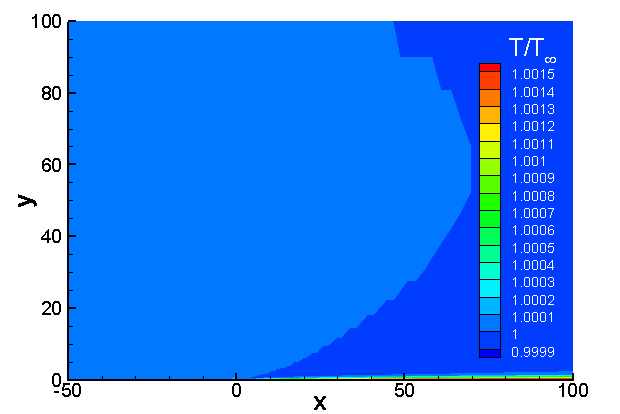}
}
\caption{\label{Fig:plate_contour} The contours of flow variables for flow over flat plate}
\end{figure}

\begin{figure}
\centering
\subfigure[\label{Fig:u5} $x=5$]{
\includegraphics[width=0.45\textwidth]{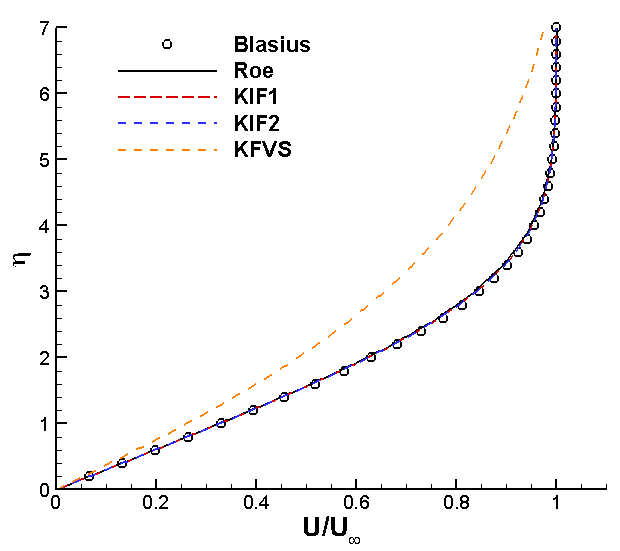}
}\hspace{0.05\textwidth}%
\subfigure[\label{Fig:u10} $x=10$]{
\includegraphics[width=0.45\textwidth]{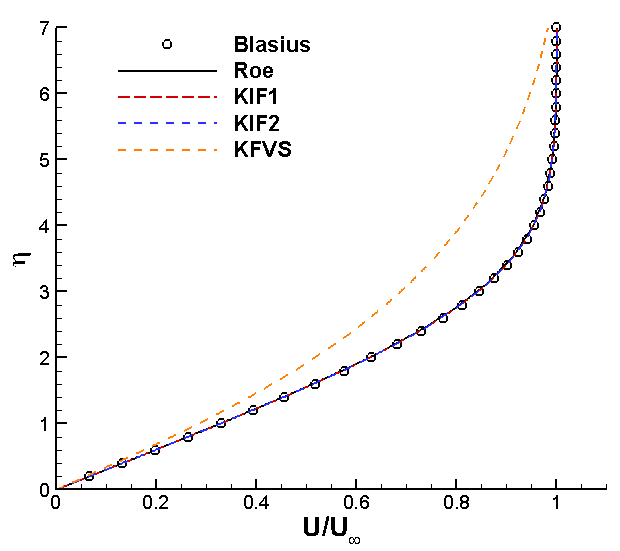}
}\\
\subfigure[\label{Fig:u20} $x=20$]{
\includegraphics[width=0.45\textwidth]{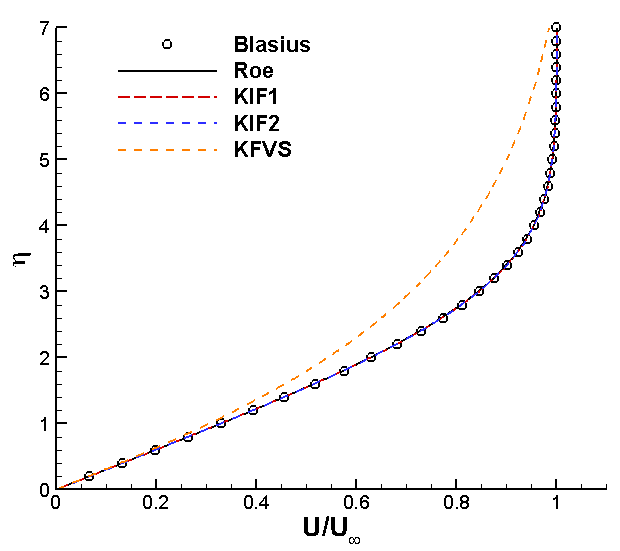}
}\hspace{0.05\textwidth}%
\subfigure[\label{Fig:u40} $x=40$]{
\includegraphics[width=0.45\textwidth]{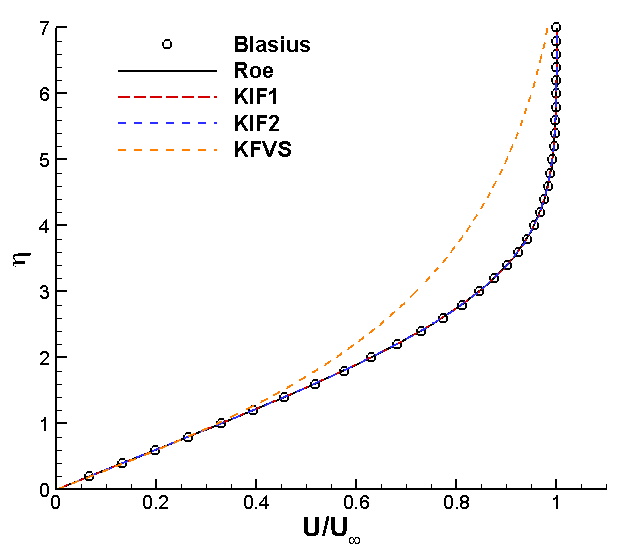}
}
\caption{\label{Fig:plate_u} The U-velocity along the vertical line at different $x$ coordinates}
\end{figure}

\begin{figure}
\centering
\subfigure[\label{Fig:v5} $x=5$]{
\includegraphics[width=0.45\textwidth]{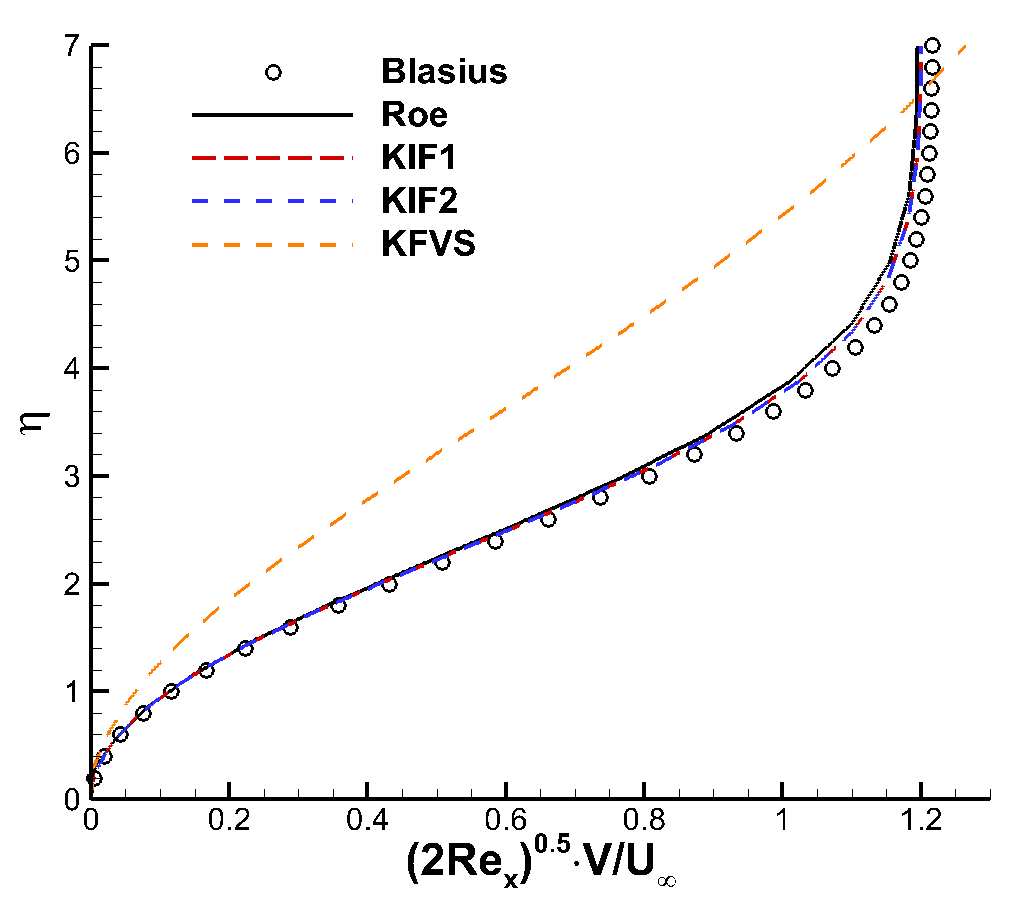}
}\hspace{0.05\textwidth}%
\subfigure[\label{Fig:v10} $x=10$]{
\includegraphics[width=0.45\textwidth]{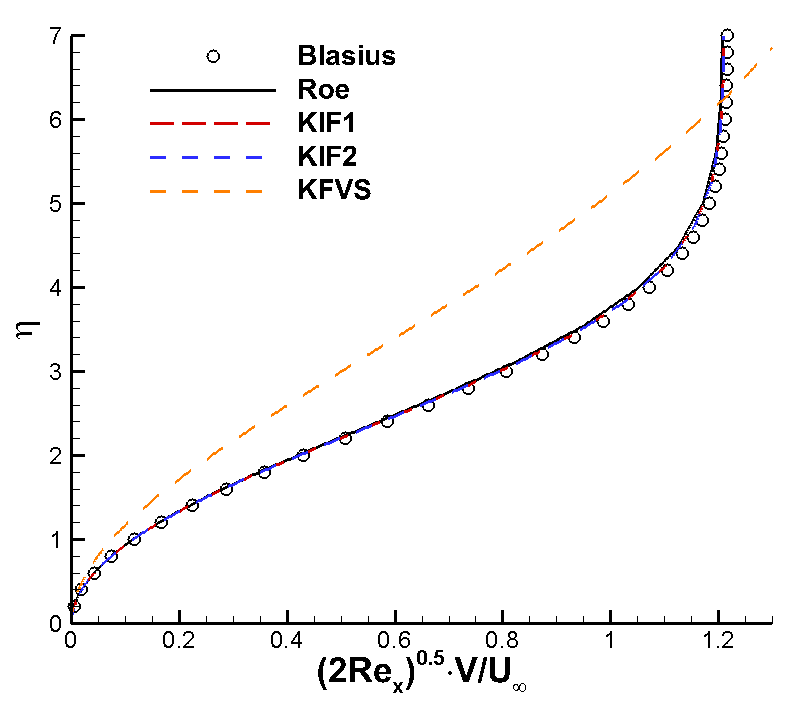}
}\\
\subfigure[\label{Fig:v20} $x=20$]{
\includegraphics[width=0.45\textwidth]{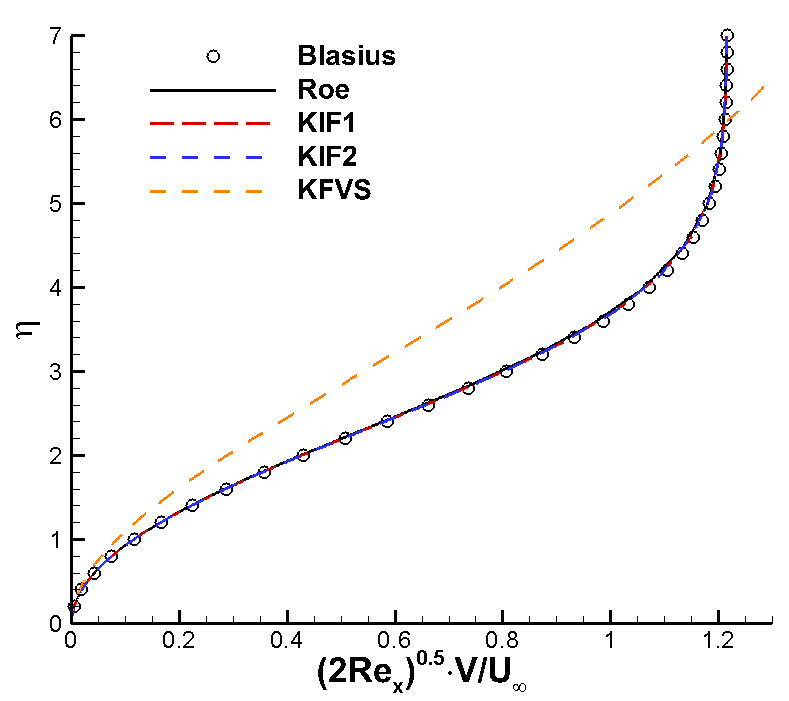}
}\hspace{0.05\textwidth}%
\subfigure[\label{Fig:v40} $x=40$]{
\includegraphics[width=0.45\textwidth]{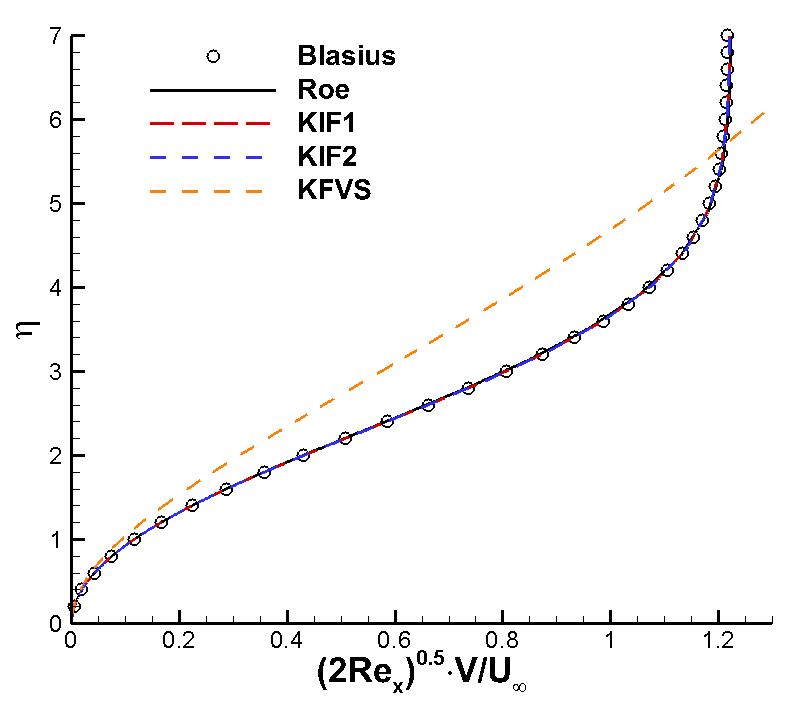}
}
\caption{\label{Fig:plate_v} The V-velocity along the vertical line at different $x$ coordinates}
\end{figure}

The flow field predicted by KIF1 is shown in Fig.~\ref{Fig:plate_contour}, the U-velocity and V-velocity along $x=5,10,20,40$ are shown in Fig.~\ref{Fig:plate_u} and Fig.~\ref{Fig:plate_v}, respectively, and are compared with the Blasius solution along with the numerical solutions from KFVS and Roe. It can be seen that the profiles of KIF1, KIF2 and Roe almost overlap with each other and the Blasuis solution. While the results of KFVS show smoother profiles and deviate much from the Blasuis solution. The reason is that it introduces extra numerical viscosity to the inviscid flux as analyzed in Sec.~\ref{Sec:KFVS}. For Both KIF1 and KIF2, since there is no discontinuity in the flow field, the weight of TTT is almost unity in this case.

\subsection{The cavity flow}
\begin{figure}
\centering
\subfigure[\label{Fig:cavity_geo} Geometry]{
\includegraphics[width=0.34\textwidth]{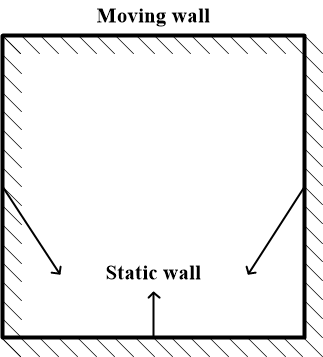}
}
\subfigure[\label{Fig:cavity_streamline} Streamline]{
\includegraphics[width=0.4\textwidth]{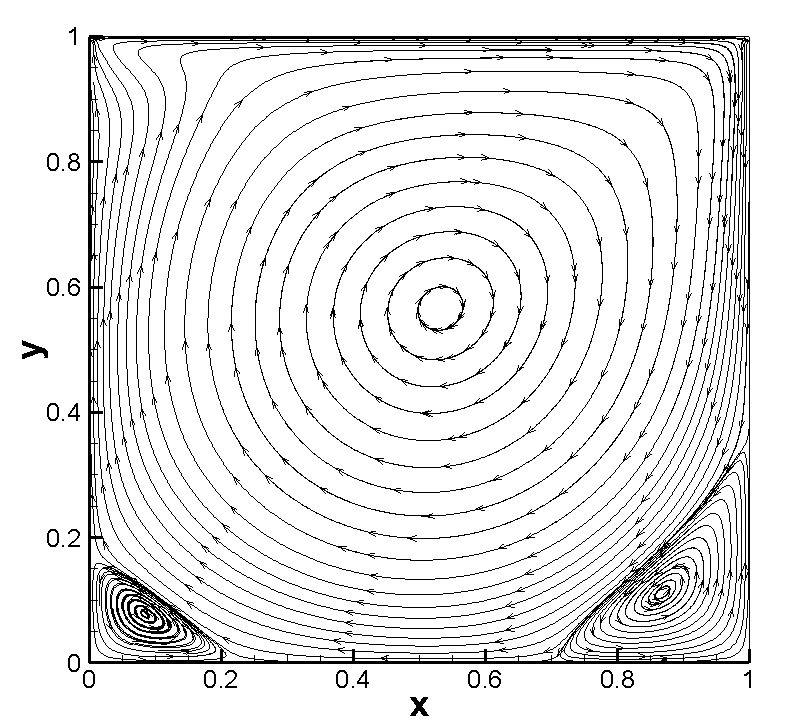}
}
\caption{\label{Fig:cavity} The geometry of the cavity flow, and the streamline calculated by KIF1}
\end{figure}

The cavity flow is a closed flow in a square region bounded by a top moving wall towards right and three static walls. The geometry with boundary conditions is shown in Fig.~\ref{Fig:cavity}. In this section, the Re$=1000$ cavity flow (the reference length is the wall length) whose top wall speed is Ma$=0.1$ is used to test the low speed property of the KIF method. The results of KIF1, KIF2 are compared with the benchmark result in Ref.~\cite{GhiaHigh}.

\begin{figure}
\centering
\subfigure[\label{Fig:cavity_u_line} U-velocity]{
\includegraphics[width=0.49\textwidth]{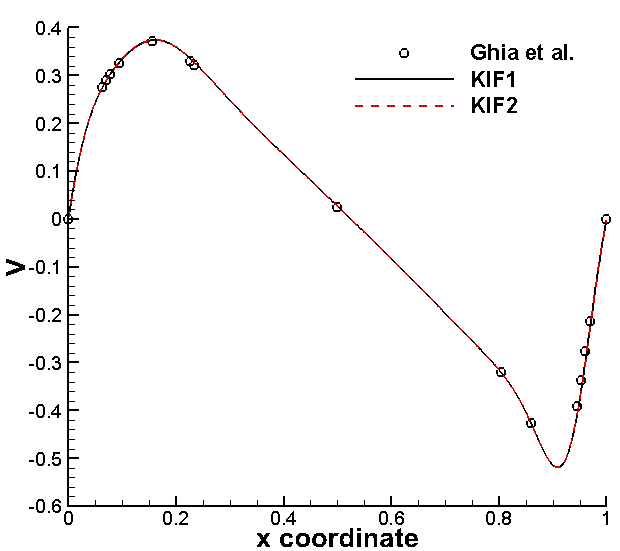}
}
\subfigure[\label{Fig:cavity_v_line} V-velocity]{
\includegraphics[width=0.49\textwidth]{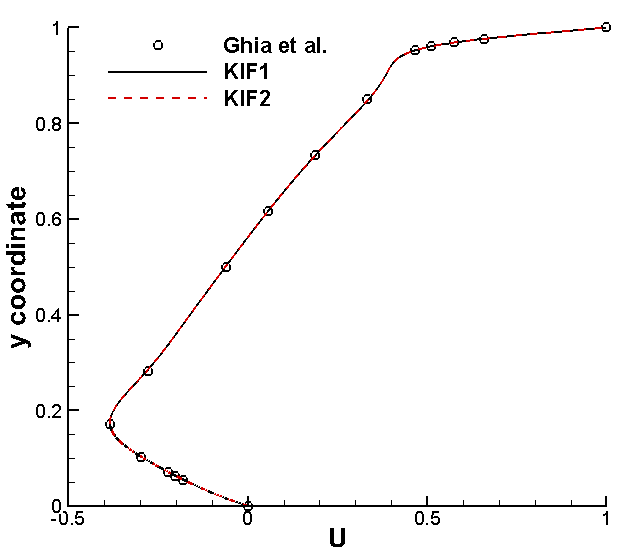}
}
\caption{\label{Fig:cavity_line} The U-velocity along the central-vertical line and the V-velocity along the central-horizontal line}
\end{figure}

The U-velocity along the central-vertical line and V-velocity along the central-horizontal line are shown in Fig.~\ref{Fig:cavity_line}, both the KIF1 and KIF2 match well with the benchmark result. Moreover, the locations of the three vortex centers in the flow field are shown in Tab.~\ref{table:vertex}, the KIF1 and KIF2 also match well with the benchmark result.

\subsection{The hypersonic cylinder flow}
The hypersonic flow passes through a circular cylinder and forms a bowl shock in front of it. The inflow Ma number is 8.03 and the inflow Re number is 183500. The reference length is the radius of the cylinder. The solid wall temperature $T_{w}$ is set 2.3566512 times of the inflow temperature $T_{\infty}$.

\begin{figure}
\centering
\includegraphics[width=0.4\textwidth]{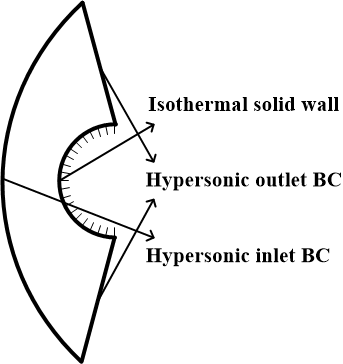}
\caption{\label{Fig:cylinder} The geometry of the hypersonic cylinder flow}
\end{figure}

The computational domain with boundary conditions is shown in Fig.~\ref{Fig:cylinder}. The supersonic inlet boundary condition and supersonic outlet boundary condition are used at the inlet and outlet, respectively. The solid wall of cylinder uses the iso-thermal boundary condition. $200\times200$ meshes are used for the whole computational domain, the height of cell adjacent to the solid wall is set $5\times10^{-5}$ to make sure that the cell Re number is not more than 10~\cite{Xu2005multidimensional}. The CFL number is set 0.8. The KIF method is compared to the AUSM+-up method which is good at hypersonic flow simulations, by imbedding the KIF into the SU2 framework~\cite{economon2016su2} (version 6.2.0), and the AUSM+-up is in this version of SU2. Therefore, the same mesh and the same reconstruction (second order MUSCL with Venkatakrishnan limiter) are adopted for both the AUSM+-up and the KIF.

\begin{figure}
\centering
\subfigure[\label{Fig:cyliner_weight1} KIF1]{
\includegraphics[width=0.29\textwidth]{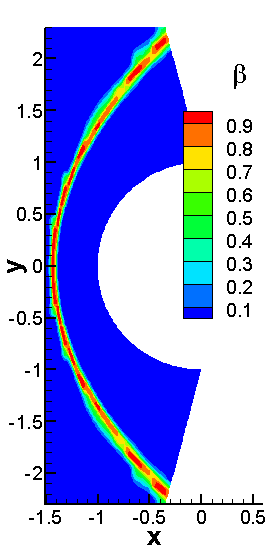}
}
\subfigure[\label{Fig:cyliner_weight2} KIF2]{
\includegraphics[width=0.29\textwidth]{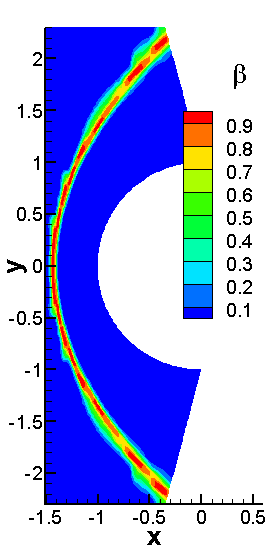}
}
\caption{\label{Fig:cyliner_weight} The contours of the weights of KFVS in the hypersonic cylinder flow}
\end{figure}

\begin{figure}
\centering
\subfigure[\label{Fig:cylinder_mach_asumplusup} AUSM+-up]{
\includegraphics[width=0.29\textwidth]{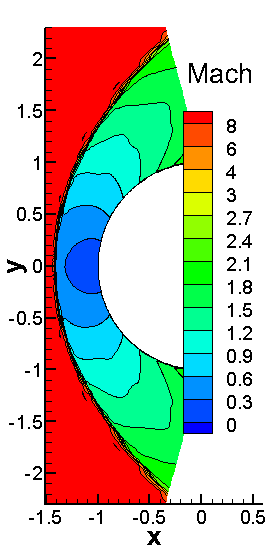}
}
\subfigure[\label{Fig:cylinder_mach_kif1} KIF1]{
\includegraphics[width=0.29\textwidth]{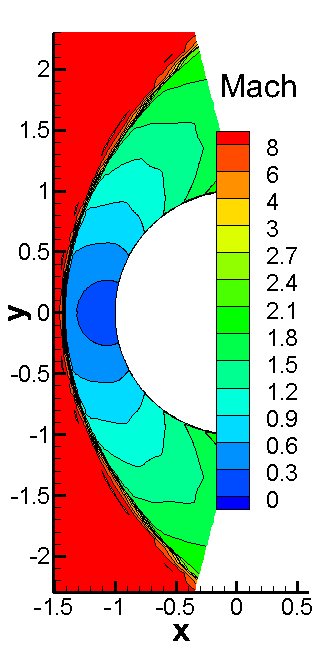}
}
\subfigure[\label{Fig:cylinder_mach_kif2} KIF2]{
\includegraphics[width=0.29\textwidth]{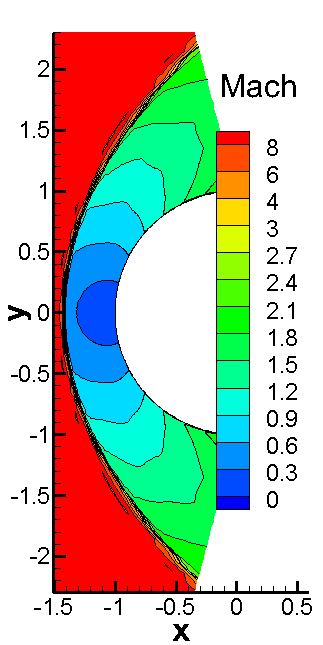}
}
\caption{\label{Fig:cylinder_mach} The mach number contours of the hypersonic cylinder flow}
\end{figure}

\begin{figure}
\centering
\subfigure[\label{Fig:cylinder_rho_asumplusup} AUSM+-up]{
\includegraphics[width=0.29\textwidth]{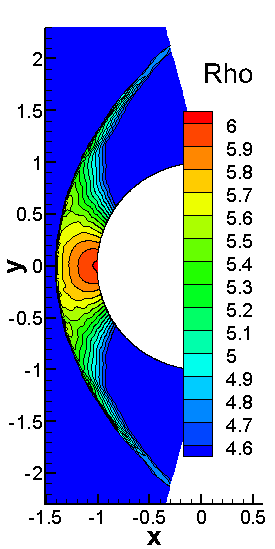}
}
\subfigure[\label{Fig:cylinder_rho_kif1} KIF1]{
\includegraphics[width=0.29\textwidth]{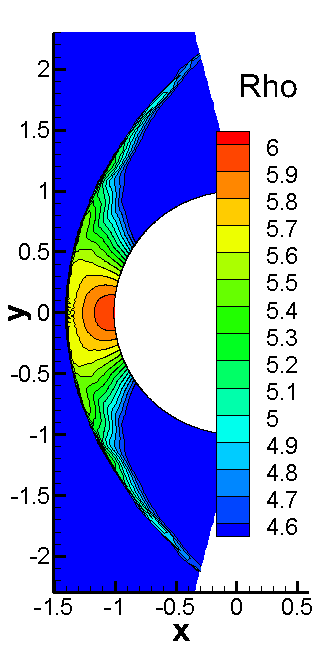}
}
\subfigure[\label{Fig:cylinder_rho_kif2} KIF2]{
\includegraphics[width=0.29\textwidth]{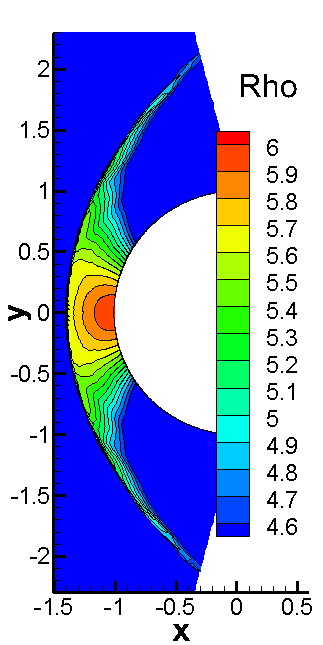}
}
\caption{\label{Fig:cylinder_rho} The density contours of the hypersonic cylinder flow}
\end{figure}

\begin{figure}
\centering
\subfigure[\label{Fig:cylinder_t_asumplusup} AUSM+-up]{
\includegraphics[width=0.29\textwidth]{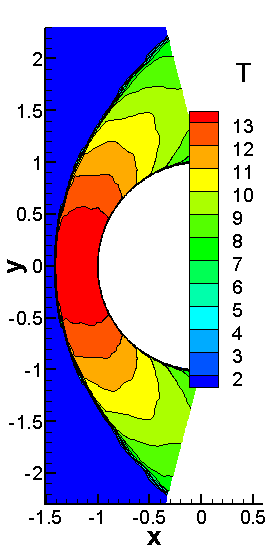}
}
\subfigure[\label{Fig:cylinder_t_kif1} KIF1]{
\includegraphics[width=0.29\textwidth]{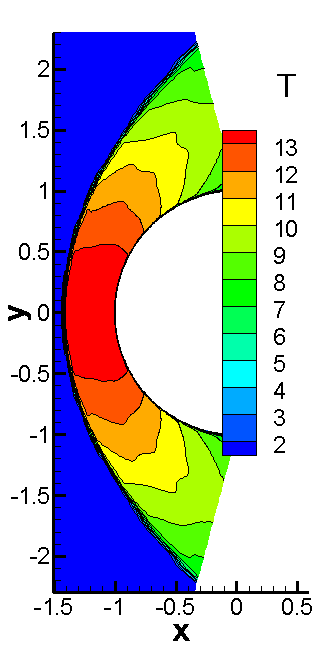}
}
\subfigure[\label{Fig:cylinder_t_kif2} KIF2]{
\includegraphics[width=0.29\textwidth]{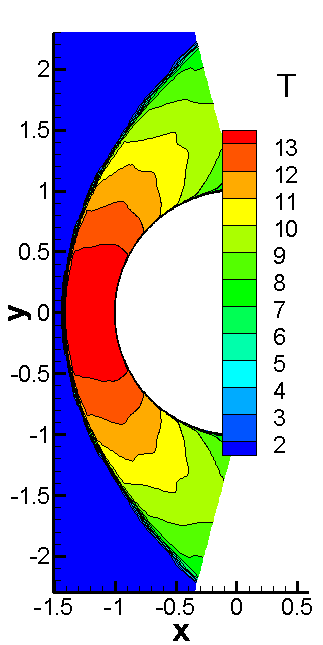}
}
\caption{\label{Fig:cylinder_t} The temperature contours of the hypersonic cylinder flow}
\end{figure}

The weight of KFVS in KIF1 and KIF2 for this case are shown in Fig.~\ref{Fig:cyliner_weight}. The KFVS only prevails near the discontinuity, while, in other regimes, TTT prevails and the weight of KFVS is almost zero. The Mach number, density and temperature contours predicted by the KIF1, KIF2 and AUSM+-up are shown in Figs.~\ref{Fig:cylinder_mach}, \ref{Fig:cylinder_rho} and \ref{Fig:cylinder_t}, respectively. Their numerical predictions match well with each other. Also, both AUSM+-up and KIF methods are not affected by the carbuncle phenomenon in this case.

For this case, due to the strong compressibility, the density contours are very sensitive. The results predicted by AUSM+-up, KIF1 and KIF2 are very similar to each other (Fig.~\ref{Fig:cylinder_rho}), except the density contour oscillates slightly after the bow-shock in AUSM+-up predictions in Fig.~\ref{Fig:cylinder_rho_asumplusup}. For both KIF1 and KIF2, the density contours are sharp and clear. The reason may be as follows: behind the shock wave and in front of the cylinder, the gradients of physical variables are very large (especially the density), numerical schemes have the trend to give oscillating results. Although the weight of KFVS in KIF is near zero, but it is not exact zero. Therefore, when some un-physical oscillations appear in the flow field and can not be resolved by the physical mesh, the small portion of KFVS takes over and treats these oscillations in meso-scale mechanism. It looks like some numerical viscosity is added to the region where un-physical oscillation occurs.

\begin{figure}
\centering
\subfigure[\label{Fig:cylinder_wall_p} Pressure]{
\includegraphics[width=0.49\textwidth]{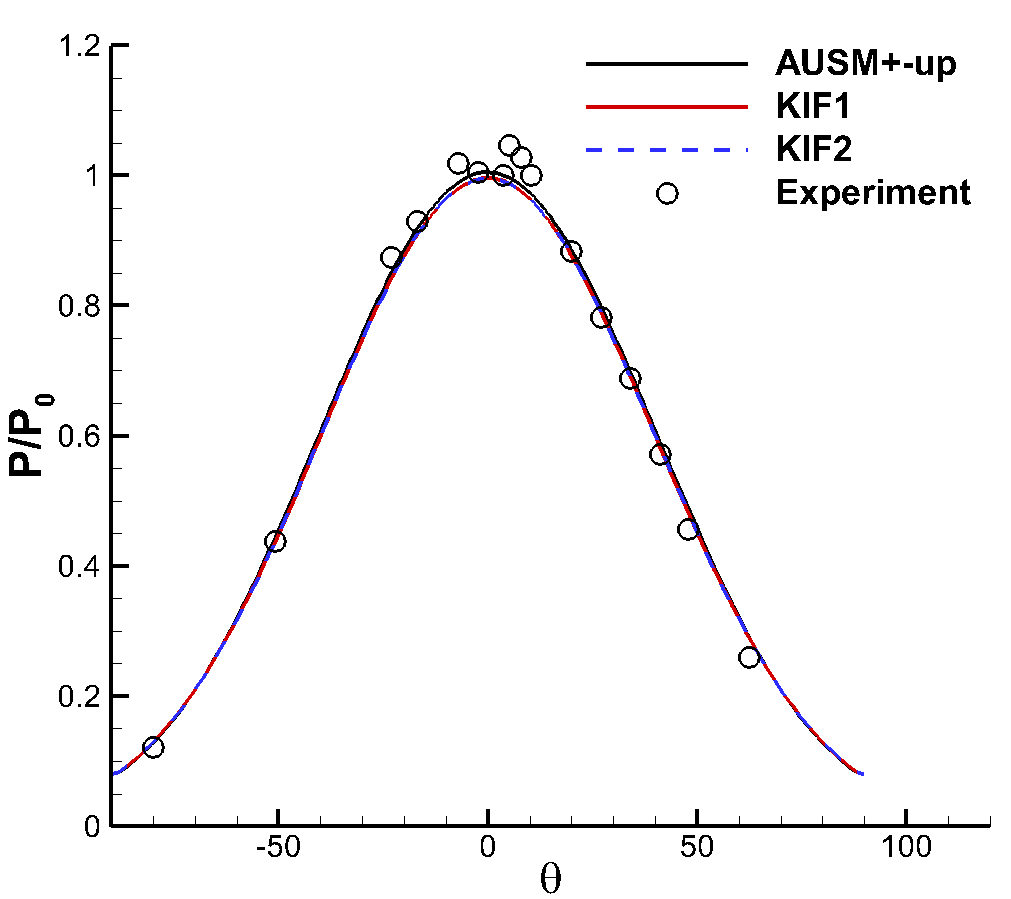}
}
\subfigure[\label{Fig:cylinder_wall_q} Heat flux]{
\includegraphics[width=0.49\textwidth]{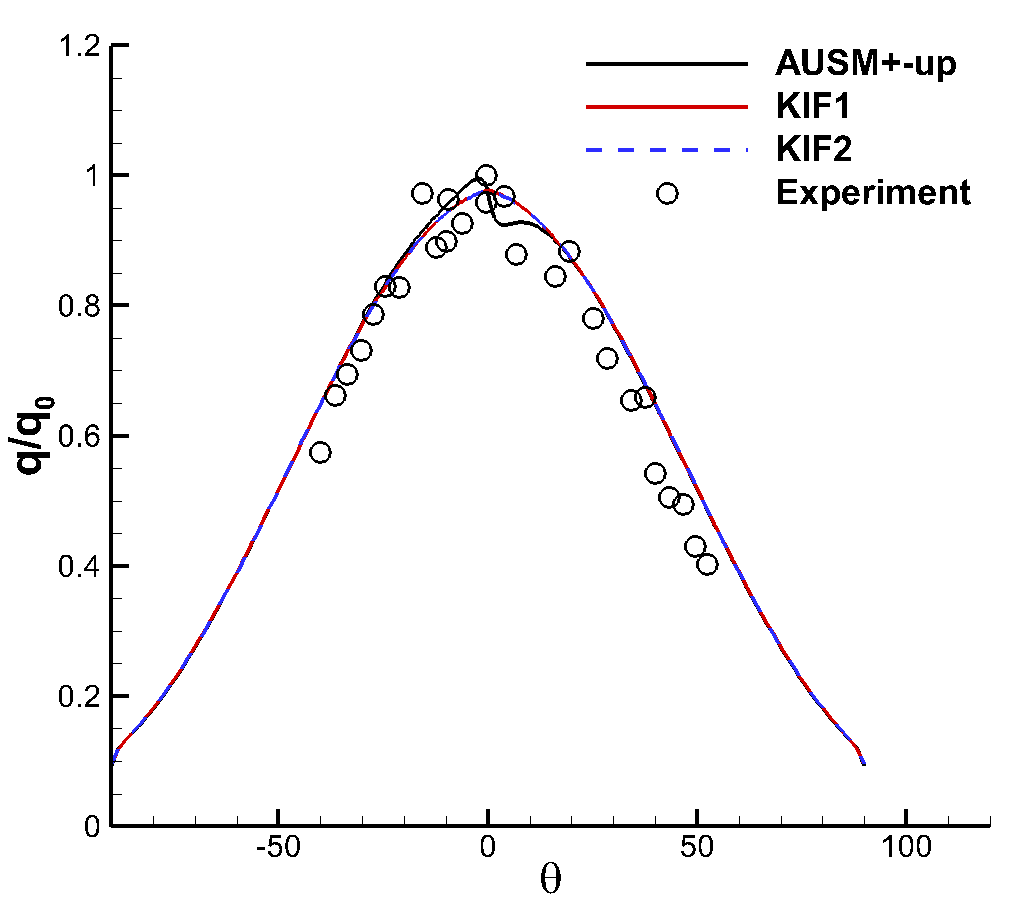}
}
\caption{\label{Fig:cylinder_wall} The pressure and heat flux on the solid wall of cylinder}
\end{figure}

For hypersonic flow, since the pressure and heat flux on solid walls are very important, the numerical results of KIF1, KIF2 and AUSM+-up are compared with experiment data in Fig.~\ref{Fig:cylinder_wall}. The pressure and heat flux are normalized by $0.9209{\rho_{\infty}}U_{\infty}^2$ and $0.003655{\rho_{\infty}}U_{\infty}^3$, respectively. It can be seen that profiles predicted by KIF1, KIF2 and AUSM+-up almost overlap with each other and match well with the experiment data, except the cusp around the stagnation point in the heat flux profile of AUSM+-up.

\section{The conclusion}\label{Sec:conclusion}
In this paper, the KIF method for solving the incompressible and compressible flows is proposed. It is can be viewed as the weighted average of the previous KFVS and TTT schemes, and the weights are determined by those in the direct modeling of multi-scale flow behavior. In the KIF, the KFVS is viewed as describing the free transport behavior of gas (in the micro-scale) with the assumption that the distribution function is the Maxwellian one, and its weight is almost unity at the discontinuity that can not be well resolved by the mesh cells; while, in smooth region, the TTT without extra numerical viscosity takes over. Two ways of choosing the weights are proposed in this paper, which are UGKS-type and DUGKS-type, respectively, and their numerical results are almost the same in the test cases of this paper. Through the comparison with the HLLC, AUSM+-up, Roe and analytical solutions, it is proved that the KIF does not suffer from the carbuncle phenomenon in high speed flow simulations and does not introduce extra numerical viscosity in low speed flow simulations (especially in the boundary layer).

\section*{Acknowledgements}
The authors thank Prof. Kun Xu at Hong Kong University of Science and Technology for discussions of the direct modeling of multi-scale flows. Junzhe Cao thanks Dr. Ji Li at Northwestern Polytechnical University for helps in imbedding flux solvers into SU2 framework. The present work is supported by National Numerical Wind-Tunnel Project of China, National Natural Science Foundation of China (Grant No. 11702223, No. 11902266 and No. 11902264) and 111 Project of China (Grant No. B17037).

\section*{Reference}
\bibliographystyle{elsarticle-num}
\bibliography{KIF}
\clearpage

\begin{table}
\centering
\caption{\label{table:vertex} The vortex centers of cavity flow.}
\begin{tabular}{p{0.2\columnwidth}<{\centering} p{0.12\columnwidth}<{\centering} p{0.12\columnwidth}<{\centering} p{0.12\columnwidth}<{\centering} p{0.12\columnwidth}<{\centering} p{0.12\columnwidth}<{\centering} p{0.12\columnwidth}<{\centering}}
    \hline
    \hline
    \multirow{2}{*}{method}
    &\multicolumn{2}{p{0.24\columnwidth}<{\centering}}{central vertex} &\multicolumn{2}{p{0.24\columnwidth}<{\centering}}{bottom-left vertex} &\multicolumn{2}{p{0.24\columnwidth}<{\centering}}{bottom-right vertex}\\
    \cmidrule(lr){2-3} \cmidrule(lr){4-5} \cmidrule(lr){6-7}
    &x &y &x &y &x &y\\
    \hline
    Ghia et al.~\cite{GhiaHigh}	&0.5313 &0.5625	&0.0859	&0.0781	&0.8594	&0.1094\\
    \hline
    KIF1	&0.5315	&0.5658	&0.0841	&0.0775	&0.8674	&0.1122\\
    \hline
    KIF2	&0.5316	&0.5658	&0.0834	&0.0776	&0.8675	&0.1123\\
    \hline
    \hline
\end{tabular}
\end{table}

\end{document}